\documentclass[aps,prb,onecolumn,showpacs]{revtex4}

\usepackage{epsfig}
\usepackage{hyperref}
\begin{document}
\draft

\def\il{I_{low}} 
\def\iu{I_{up}}
\def\eeq{\end{equation}}
\def\ie{i.e.}  
\def\etal{{\it et al. }}  
\def\prb{Phys. Rev. {B }}
\def\epjb{Eur. Phys. J. {B }}
\def\pra{Phys. Rev. {A}} 
\def\prl{Phys. Rev. Lett. }
\def\pla{Phys. Lett. A } 
\def\pb{Physica B}
\def\pt{Physics Today }
\def\ajp{Am. J. Phys. }  
\def\mpl{Mod. Phys. Lett. { B}} 
\def\ijmp{Int. J. Mod. Phys. { B}} 
\def\ijp{Ind. J. Phys. }
\def\ijpap{Ind. J. Pure Appl. Phys. }
\def\ibmjrd{IBM J. Res. Dev. }
\def\pjp{Pramana J. Phys.}
\def\ltp{J. of Low Temp. Phys.}
\def\jpcm{J. of Phys. Condensed Matter}

\title{Resolving the order parameter of High-T$_{c}$ Superconductors
  through quantum pumping spectroscopy} 
\author{Colin Benjamin}
\email[E-Mail me at:  colin\_ben@indiatimes.com  ]{  or at, colin@sa.infn.it}
\homepage[Homepage: ] {http://www.iopb.res.in/~colin}
\altaffiliation[Present Address: ]{Dipartimento di Fisica "E. R. Caianiello",
Universita  degli Studi di Salerno,
Via S. Allende,
I-84081 Baronissi (SA),
Italy}
\affiliation{Institute of Physics,
  Sachivalaya Marg, Bhubaneswar 751 005, Orissa, India}

\date{\today} 

\begin{abstract}
  The order parameter of High-T$_{c}$ superconductors through a series
  of experiments has been quite conclusively demonstrated to not be of
  the normal $s-wave$ type. It is either a pure $d_{x^{2}-y^{2}}$-wave
  type or a mixture of a $d_{x^{2}-y^{2}}-wave$ with a small imaginary
  $s-wave$ or $d_{xy}-wave$ component. In this work a distinction is brought out among the four types, i.e., $s- wave$, $d_{x^{2}-y^{2}}- wave$,
  $d_{x^{2}-y^{2}}+is - wave$ and $d_{x^{2}-y^{2}}+id_{xy}- wave$ types with the help of quantum pumping
  spectroscopy. This involves a normal metal double barrier structure
  in contact with a High-T$_{c}$ superconductor. The pumped current,
  heat and noise show different characteristics with change in order
  parameter revealing quite easily the differences among these.

\end{abstract}

\maketitle 
\section {Introduction}

One of the outstanding issues of High-T$_{c}$ superconductor research
involves the identification of the order parameter symmetry and the
underlying mechanism\cite{bruder,rice}. Although a host of experiments
have indicated the order parameter symmetry to be of a
$d_{x^{2}-y^{2}}-wave$ type\cite{vanhar,tseui}, there are
theoretical works\cite{stefankis,asano1,asano2,shiba} which indicate
that an imaginary $s-wave$ or $d_{xy}-wave$ component is necessary to explain some of
the experimental results. These experimental results\cite{kirtley}
being notably the splitting of the zero energy peak in conductance
spectra which indicates the presence of an imaginary $s-wave$ or $d_{xy}-wave$ 
component which would break the time reversal symmetry. Many
theoretical attempts have been made to bring out the differences among
the different order parameters. The first theoretical attempts were
made by Hu in Ref.[\onlinecite{hu}] where the existence of a sizable
areal density of midgap states on the \{110\} surface of a
$d_{x^{2}-y^{2}}-wave$ superconductor was brought out. Further using
tunneling spectroscopy, Tanaka and Kashiwaya in
Ref.[\onlinecite{tanaka}] brought out the fact that zero bias
conductance peaks (which were seen earlier in many
experiments\cite{becherer}) are formed when a normal metal is in
contact with a $d_{x^{2}-y^{2}}-wave$ superconductor enabling a
distinction between $s-wave$ and $d_{x^{2}-y^{2}}-wave$
superconductors. A shot noise analysis by Zhu and Ting in
Ref.[\onlinecite{zhu}] also revealed differences between $s-wave$ and
$d_{x^{2}-y^{2}}-wave$ superconductors. Further inclusion of phase
breaking effects\cite{belogovskii} in double barriers formed by normal
and superconducting electrodes revealed a double peaked structure in
case of $s-wave$ while a dramatic reduction of zero bias maximum for
$d_{x^{2}-y^{2}}-wave$ superconductors. These are in addition to many
other works which involve spin polarized transport in
ferromagnet-superconductor junctions\cite{zutic,hirai,kashiwaya} which
reveal differences between different possible High-T$_c$ order
parameters. In a recent review, Deutscher\cite{deutscher} has used the
Andreev-saint James reflections to indicate the presence of an
additional imaginary component in the order parameter.  Also in
another review\cite{lofwander}, Lofwander, et. al., arrived at some
conclusive arrivals for $d_{x^{2}-y^{2}}-wave$ superconductivity in
the cuprates. Recently, Ng and Varma\cite{varma} studied some of the
proposed order parameters and also suggested new experiments to bring
out the subtle differences among these.  In this work we apply the
principles of quantum adiabatic pumping to bring out the differences
between the different types of order parameters. Quantum adiabatic
pumping involves the transport of particles without the application of
any bias voltage.  This is done by varying in time atleast two
independent parameters of the mesoscopic system out of phase. The
physics of the adiabatic quantum pump is based on two independent
works by Brouwer in Ref.[\onlinecite{brouwer}] and by Zhou, et.  al.,
in Ref.[\onlinecite{zhou}] which built on earlier works by BTP in
Ref.[\onlinecite{bpt}]. The first experimental realization of an
adiabatic quantum pump was made in Ref.[\onlinecite{switkes}]. The
phenomenon of quantum adiabatic pumping has been extended to pump a
spin current\cite{spin_wu_wang} also it has been used in different
mesoscopic systems like quantum hall systems\cite{blau_1}, luttinger
liquid based mesoscopic conductor\cite{citro}, in the context of
quantized charge pumping due to surface acoustic waves\cite{aharony},
a quantum dot in the kondo regime\cite{wang_kondo}, and of course in
the context of enhanced pumped currents in hybrid mesoscopic systems
involving a superconductor\cite{wang_apl,blaau}. In
Ref.[\onlinecite{wang_apl}], Jian Wang, et. al., showed that andreev
reflection at the junction between a normal metal and a superconductor
(of, $s-wave$ type) can enhance the pumped current as much as four
times that in a purely normal metal structure.  M. Blauuboer in
Ref.[\onlinecite{blaau}] showed that for slightly asymmetric coupling
to the leads, this enhancement can be slightly increased. Recently,
Taddei, et.al. in Ref.[\onlinecite{taddei}], generalized the adiabatic
quantum pumping mechanism wherein several superconducting leads are
present.

This work is organized as follows- After generalizing the formula for
the adiabatically pumped current through a normal metal lead in
presence of a High-T$_c$ superconductor, we derive the amount of
pumped charge current into the normal metal in the vicinity of a
High-T$_c$ superconductor with different types of order parameter
symmetry. Next we focus on the heat transported and noise generated in
the pumping process in case of each of the specific order parameter
symmetries. Finally we juxtapose all the obtained results in case of
different order parameter symmetry in the amount of pumped current,
heat and noise to have some conclusive arrivals and to propose
experiments which would fulfill this theoretical proposal.

\begin{figure}[h]
  \protect\centerline{\epsfxsize=5.2in\epsfbox{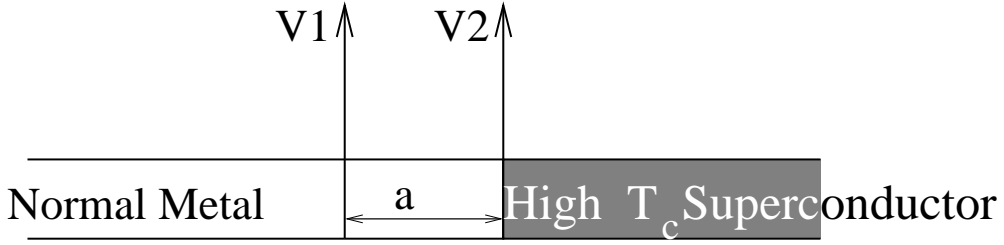}}
\caption{The model system. A normal metal double barrier structure in proximity with a High $T_{c}$ superconductor. The double barrier structure is modeled by two delta barriers distance $a$ apart.}
\end{figure}

\section {Theory of the pumped charge current}

The model system is shown in Fig.~1. It consists of a normal metal
double barrier structure in junction with a High-T$_c$ superconductor.
The double barrier structure is modeled by two delta barrier's of
strengths $V_1$ and $V_2$, a distance '$a$' apart. Quantum pumping is
enabled by adiabatic modulations in the strength of the delta
barriers, i.e., $V_{1}=V_{0}+V_{p}sin(wt)$ and
$V_{2}=V_{0}+V_{p}sin(wt+\phi)$, where $V_p$ is the strength of the
pumping amplitude.  Andreev reflection
mechanism\cite{andreev,beenakker_rmp} is what takes place when a
normal metal is brought in contact with a superconductor. The
scattering matrix for the entire system is given by:

\begin{equation}
S_{NS}(\epsilon)=\left(\begin{array}{cc}
S_{ee}(\epsilon)       & S_{eh}(\epsilon) \\
S_{he}(\epsilon)  &     S_{hh}(\epsilon)  \\
\end{array} \right) 
\end{equation}

wherein
$S_{ee}(\epsilon),S_{eh}(\epsilon),S_{he}(\epsilon),S_{hh}(\epsilon)$
are 1X1 matrices, since we are considering single channel leads. The
explicit analytical form of the expressions are given
by\cite{beenakker}:

\begin{eqnarray}
S_{ee}(\epsilon)&=&S_{11}(\epsilon)+\frac{S_{12}(\epsilon)\alpha^{h}S^{*}_{22}(-\epsilon)\alpha^{e}S_{21}(\epsilon)}{1-\alpha^{h}\alpha^{e}S_{22}(\epsilon)S^{*}_{22}(-\epsilon)},\nonumber\\
S_{he}(\epsilon)&=&\frac{S^{*}_{12}(-\epsilon)\alpha^{e}S^{*}_{21}(\epsilon)}{1-\alpha^{h}\alpha^{e}S_{22}(\epsilon)S^{*}_{22}(-\epsilon)},\nonumber\\
S_{eh}(\epsilon)&=&\frac{S^{}_{12}(\epsilon)\alpha^{h}S^{*}_{21}(-\epsilon)}{1-\alpha^{h}\alpha^{e}S_{22}(\epsilon)S^{*}_{22}(-\epsilon)},\nonumber\\
S_{hh}(\epsilon)&=&S^{*}_{11}(-\epsilon)+\frac{S^{*}_{12}(-\epsilon)\alpha^{e}S^{}_{22}(\epsilon)\alpha^{h}S^{*}_{21}(-\epsilon)}{1-\alpha^{h}\alpha^{e}S_{22}(\epsilon)S^{*}_{22}(-\epsilon)}.\\
\mbox{ with, } \alpha^{h}&=&e^{-i \mbox{ arccos} [\frac{\epsilon}{\Delta(k_{h})}]+i\phi(k_{h})},  \alpha^{e}=e^{-i \mbox{ arccos} [\frac{\epsilon}{\Delta(k_{e})}]-i\phi(k_{e})},\nonumber\\
e^{i\phi(k_{e})}&=& \frac{\Delta(k_{e})}{|\Delta(k_{e})|}, \mbox{ and  } e^{i\phi(k_{h})}=\mbox{} \frac{\Delta(k_{h})}{|\Delta(k_{h})|}.
\end{eqnarray}

where, $\phi(k_{e})$ and $\phi(k_{h})$ are the phase of the order
parameter for electronic like quasiparticles and hole like
quasiparticles respectively, with $k_e$ and $k_h$ being the respective
wavevectors for the electronic like quasiparticles and hole like
quasiparticles\cite{belogovskii}.

  From Refs.[\onlinecite{wang_apl,blaau}], the adiabatically pumped electronic current into the normal lead in presence of the High-T$_c$ superconducting
  lead is given by-

\begin{equation}
I_{e}=\frac{wq_{e}}{2\pi}\int_{0}^{\tau} d\tau[\frac{dN^{e}_{L}}{dV_{1}} \frac{dV_{1}}{dt}       + \frac{dN^{e}_{L}}{dV_{2}}  \frac{dV_{2}}{dt}],
\end{equation}

The quantity $dN^{e}_{L}/dV$ (wherein, the subscript $_L$ denotes left
lead or the normal lead) is the electronic injectivity given at zero temperature by
\begin{equation}
\frac{dN^{e}_{L}}{dV_{j}}=\frac{1}{2\pi} \Im[S^{*}_{ee}\partial_{V_{j}}S_{ee}+S^{*}_{eh}\partial_{V_{j}}S_{eh}]
\end{equation}

In the above equation and below, $\Im $ represents the imaginary part of the quantity in parenthesis. Similarly, the adiabatically pumped hole current into the normal lead in presence of the High-T$_c$ superconducting  lead is given by-

\begin{equation}
I_{h}=\frac{wq_{h}}{2\pi}\int_{0}^{\tau} d\tau[\frac{dN^{h}_{L}}{dV_{1}} \frac{dV_{1}}{dt}       + \frac{dN^{h}_{L}}{dV_{2}}  \frac{dV_{2}}{dt}],
\end{equation}

The quantity $dN^{h}_{L}/dV$ (wherein, the subscript $_L$ denotes left
lead or the normal lead) is the hole injectivity given at zero temperature by
\begin{equation}
\frac{dN^{h}_{L}}{dV_{j}}=\frac{1}{2\pi} \Im[S^{*}_{hh}\partial_{V_{j}}S_{hh}+S^{*}_{he}\partial_{V_{j}}S_{he}]
\end{equation}
with $q_{e}=-q_{h}$ as per the usual convention,
and in the weak pumping regime the adiabatically pumped electronic current similar to the analysis in Refs.[\onlinecite{brouwer,wang_apl}], is given
by\cite{wang_comm}-
\begin{equation}
I_{e}=\frac{wq_{e}sin(\phi)V^{2}_{p}}{\pi} \Im[\partial_{V_{1}}S^{*}_{ee}\partial_{V_{2}}S_{ee}+\partial_{V_{1}}S^{*}_{eh}\partial_{V_{2}}S_{eh}]
\end{equation}

 and the adiabatically pumped hole current in the weak pumping regime is 

\begin{equation}
I_{h}=\frac{wq_{h}sin(\phi)V^{2}_{p}}{\pi} \Im[\partial_{V_{1}}S^{*}_{hh}\partial_{V_{2}}S_{hh}+\partial_{V_{1}}S^{*}_{he}\partial_{V_{2}}S_{he}]
\end{equation}

while for a normal metal structure, the expression for the pumped electronic 
current in the weak pumping regime is given by-
\begin{equation}
I_{N}=\frac{wq_{e}sin(\phi)V^{2}_{p}}{\pi} \Im[\partial_{V_{1}}S^{*}_{11}\partial_{V_{2}}S_{11}+\partial_{V_{1}}S^{*}_{21}\partial_{V_{2}}S_{21}]
\end{equation}

\section{Pumped current for different order parameters}

In Ref.[\onlinecite{wang_apl}], the pumped current for a NS system
(where the superconductor is of $s-wave$ type) has been shown to be
four times of that in a purely normal metal junction. The system
considered in Ref.[\onlinecite{wang_apl}] is also a double
delta barrier structure. We re-derive the results for the pumped current in a
normal metal-$s-wave$ superconductor junction and subsequently derive
the results for the pumped current in a normal metal-$d_{x^{2}-y^{2}}$
wave superconductor junction, for the pumped current in a normal
metal - $d_{x^{2}-y^{2}}+is$ wave superconducting junction and finally for the  pumped current in a normal metal - $d_{x^{2}-y^{2}}+id_{xy}$ wave superconducting junction. We consider in the examples below as well as in the succeding sections the fermi energy to match the chemical potential of the superconducting lead so that $\epsilon=0$. In which case, from Eq.~2  we have $S_{ee}=S^{*}_{hh}$. The system we consider is a normal metal double barrier structure at resonance in junction with a High-T$_c$ superconductor. The resonance condition in the normal metal quantum dot structure is exemplified by the fact that the reflection coefficients are zero while the transmission coefficients are unity. Thus, $|S_{11}|^{2}=|S_{22}|^{2}=0$, while $|S_{12}|^{2}=|S_{21}|^{2}=1$, with $S_{12}=S_{21}=e^{-2ika}$ for the double barrier quantum dot at resonance. Further from Eq.~2, we have, $S_{eh}=\alpha^{h}$ and $S_{he}=\alpha^{e}$, with this we get $\partial_{V_j}S_{eh}=\partial_{V_j}S_{he}=0, \mbox { for } j=1,2.$ Thus, by the arguments above the pumped electron and hole currents are exactly one and the same in both magnitude as well as direction and reduce to-
\begin{equation}
I_{h}=I_{e}=\frac{wq_{e}sin(\phi)V^{2}_{p}}{\pi} \Im[\partial_{V_{1}}S^{*}_{ee}\partial_{V_{2}}S_{ee}]
\end{equation}

Further for the double barrier structure at resonance from Eq.~2, one has the normal scattering amplitude, $S_{ee}=S_{11}+\alpha^{h}\alpha^{e}(S_{12})^{2}S^{*}_{22}$, and for the partial derivatives appering in Eq. 11, we have $\partial_{V_{1}}S_{ee}=\partial_{V_{1}}S_{11}+\alpha^{h}\alpha^{e}(S_{12})^{2}\partial_{V_{1}}S^{*}_{22}$, with the help of the dyson equation, $\partial_{V_{j}}G^{r}_{\alpha\beta}=G^{r}_{\alpha j}G^{r}_{j \beta}$, and the Fisher-Lee relation, $S_{\alpha\beta}=-\delta_{\alpha\beta}+i 2k G^{r}_{\alpha\beta}$, one can easily derive $\partial_{V_1}S_{11}=\frac{-i}{2k}$, and $\partial_{V_1}S_{22}=\frac{-i}{2k}(S_{12})^{2}$. Thus for a double barrier quantum dot at resonance, we have for the partial derivatives appearing in Eq.~11,
 
\begin{equation}
\partial_{V_1}S_{ee}=\frac{-i}{2k}(1-\alpha^{h}\alpha^{e}), \mbox { and }\partial_{V_2}S_{ee}=\frac{-i}{2k}e^{-4ika}(1-\alpha^{h}\alpha^{e}).
\end{equation}

With these formulas in mind we herein below derive the results for the pumped charge current for a normal metal double barrier structure in junction with a High-T$_c$ superconductor, which we assume to have $d_{x^{2}-y^{2}}-wave$, $d_{x^{2}-y^{2}}+is-wave$ and $d_{x^{2}-y^{2}}+id_{xy}-wave$ order parameters. For the sake of completeness and comparison we rederive the already known results for  a pure normal metal structure and that of a normal metal double barrier structure in junction with an isotropic $s-wave$ superconductor.

{ \underline {$A \mbox {  } pure\mbox {  } normal\mbox {  } metal \mbox {  }double\mbox {  } barrier\mbox {  } structure$:}}  From the discussion above the pumped current in case of a normal metal double barrier structure at resonance reduces to (from Eq.10)-
\begin{equation}
I_{N}=\frac{-wq_{e}sin(\phi)V^{2}_{p}}{4\pi k^{2}} sin (4ka)
\end{equation}

{ \underline {$An\mbox {  } isotropic\mbox {  } s-wave \mbox{ } superconductor$:}} For a normal
$s-wave$ superconductor which is isotropic
$\Delta(k_{h})=\Delta(k_{e})=\Delta$ and $\alpha^{h}=\alpha^{e}=-i$.
Thus, $\partial_{V_1}S_{ee}=\frac{-i}{k}$, and  $\partial_{V_2}S_{ee}=\frac{-i}{k}e^{-4ika}$,
 and therefore in the weak
pumping regime for an isotropic $s-wave$ superconductor in junction
with a normal metal double barrier heterostructure the pumped current
denoted by $I(NS)$ is four times that in a pure normal metal
structure\cite{wang_apl},

\begin{equation}
I(NS)= 4 I(N),\mbox { with $I_{N}$ as given in Eq. 13.} 
\end{equation}

{\underline { $d_{x^{2}-y^{2}}- wave \mbox{ } superconductor$}:} Now
we consider the case of a $d_{x^{2}-y^{2}}-wave$ superconductor, in
junction with a normal metal double barrier structure at resonance.
The effective order parameter of the $d_{x^{2}-y^{2}}-wave$
superconductor for electron like quasiparticles is
$\Delta(k_{e})=\Delta_{d}cos(2\theta_{s}-2\alpha)$ and for hole like
quasiparticles it is $\Delta(k_{h})=\Delta_{d}cos(2\theta_{s}+2\alpha)$,
with  $\theta_{s}$ being the injection angle between the
electron wave vector($k_{e}$) and the x-axis, while $\alpha$ is the
misorientation angle between the $a$ axis of the crystal and the
interface normal. Now for a $d_{x^{2}-y^{2}} -wave $ superconductor with a ($110$) orientation, $\alpha={\pi}/{4}$. Thus, $\Delta(k_{e})=\Delta_{d}sin(2\theta_{s})$ and $\Delta(k_{h})=-\Delta_{d}sin(2\theta_{s})$. In light of this we have, $e^{i\phi(k_{e})}=1$, and   $e^{i\phi(k_{h})}=-1$ and thus $\alpha_{e}=-i$ and $\alpha_{h}=i$, therefore we have $\partial_{V_1}S_{ee}=\partial_{V_2}S_{ee}=0$, and hence in the weak
pumping regime for a $d_{x^{2}-y^{2}}-wave$ superconductor in junction
with a normal metal double barrier heterostructure the pumped current
denoted by $I(ND)$ regardless of the injection angle is zero. 

\begin{equation}
I(ND)=0
\end{equation}

{\underline {$d_{x^{2}-y^{2}}+is -wave \mbox{ } superconductor$}:}
Now, we consider the order parameter of the High-T$_{c}$
superconductor to be a mixture of the $d_{x^{2}-y^{2}}+is$ type. The
$d_{x^{2}-y^{2}}$ component has a $(110)$ oriented surface, with
$\alpha=\frac{\pi}{4}$. The effective order parameter for electron and hole like quasi-particles becomes:
\begin{eqnarray*}
\Delta(k_{e})=\Delta_{d}sin(2\theta_{s})+i\Delta_{s},\mbox {and } \Delta(k_{h})=-\Delta_{d}sin(2\theta_{s})+i\Delta_{s}.
\end{eqnarray*}
For the phases of the pairing symmetries for electron and hole like quasiparticles,  we have- 
\[e^{i\phi(k_{e})}=\frac{\Delta_{d}sin(2\theta_{s})+i\Delta_{s}}{\sqrt{\Delta_{d}^{2}sin^{2}(2\theta_{s})+\Delta^{2}_{s}}}, \mbox{ and }
e^{i\phi(k_{h})}=\frac{-\Delta_{d}sin(2\theta_{s})+i\Delta_{s}}{\sqrt{\Delta_{d}^{2}sin^{2}(2\theta_{s})+\Delta^{2}_{s}}}
\]

and hence, the product $\alpha^{h}\alpha^{e}$ reduces to-
\begin{eqnarray*}
\alpha^{h}\alpha^{e}=\frac{\Delta_{d}sin(2\theta_{s})-i\Delta_{s}}{\Delta_{d}sin(2\theta_{s})+i\Delta_{s}}
\end{eqnarray*}

and finally for the partial derivatives appearing in Eq.~11 one has-
\begin{equation}
\partial_{V_{1}}S_{ee}=\frac{\Delta_{s}}{k(\Delta_{d}sin(2\theta_{s})+i\Delta_{s})}\mbox { and } \partial_{V_{2}}S_{ee}=\frac{\Delta_{s}e^{-4ika}}{k(\Delta_{d}sin(2\theta_{s})+i\Delta_{s})}
\end{equation}
Thus, the pumped charge current reduces to:
\begin{eqnarray}
I(NDs)&=&\frac{-wq_{e}sin(\phi)V^{2}_{p}}{\pi k^{2}} \frac{\Delta^{2}_{s}}{\Delta^{2}_{d}sin^{2}(2\theta_{s})+\Delta^{2}_{s}} sin (4ka)
\end{eqnarray}

From Eq.~13 and Eq.~17, the ratio of the pumped current in presence of the High-T$_{c}$
superconductor to that in a pure normal metal double barrier structure becomes:
\begin{equation}
\frac{I(NDs)}{I(N)}=4\frac{\Delta^{2}_{s}}{\Delta^{2}_{d}sin^{2}(2\theta_{s})+\Delta^{2}_{s}}
\end{equation}

 From the expression it is evident that the maximum enhancement of the pumped current is 4 times of that in a pure normal metal structure. Depending on the relative magnitudes of $\Delta_{s}$ and  $\Delta_{d}$ and the injection angle $\theta_{s}$ the ratio  ${I(NDs)}/{I(N)}$ can be as low as zero as in the pure $d_{x^{2}-y^{2}}$ case or as large as $4$ as in the pure $s-wave$ case.

{\underline {$d_{x^{2}-y^{2}}+id_{xy} -wave \mbox{ } superconductor$}:}
Finally, we consider the order parameter of the High-T$_{c}$
superconductor to be a mixture of the $d_{x^{2}-y^{2}}+id_{xy}$ type. The order parameter for electron like quasi particles is: $\Delta(k_{e})=\Delta_{d}cos(2\theta_{s}-2\alpha)+i\Delta^{'}_{d}sin(2\theta_{s}-2\alpha)$ while for hole like quasiparticles it becomes: $\Delta(k_{h})=\Delta_{d}cos(2\theta_{s}+2\alpha)+i\Delta^{'}_{d}sin(2\theta_{s}-2\alpha)$.
The $d_{x^{2}-y^{2}}$ and $d_{xy}$ component have a $(110)$ oriented surface, with $\alpha=\frac{\pi}{4}$.
Thus,
\begin{eqnarray*}
\Delta(k_{e})=\Delta_{d}sin(2\theta_{s})-i\Delta^{'}_{d}cos(2\theta_{s}),\mbox {and } \Delta(k_{h})=-\Delta_{d}sin(2\theta_{s})-i\Delta^{'}_{d}cos(2\theta_{s}).
\end{eqnarray*}

For the phases of the order parameter for electron and hole like quasi-particles we have:
\[e^{i\phi(k_{e})}=\frac{\Delta_{d}sin(2\theta_{s})-i\Delta^{'}_{d}cos(2\theta_{s})}{\sqrt{\Delta_{d}^{2}sin^{2}(2\theta_{s})+\Delta^{'2}_{d}cos^{2}(2\theta_{s})}}, \mbox{ and }
e^{i\phi(k_{h})}=\frac{-\Delta_{d}sin(2\theta_{s})-i\Delta^{'}_{d}cos(2\theta_{s})}{\sqrt{\Delta_{d}^{2}sin^{2}(2\theta_{s})+\Delta^{'2}_{d}cos^{2}(2\theta_{s})}}
\]

and hence the product $\alpha^{h}\alpha^{e}$ reduces to:
\begin{eqnarray*}
\alpha^{h}\alpha^{e}=\frac{\Delta_{d}sin(2\theta_{s})+i\Delta^{'}_{d}cos(2\theta_{s})}{\Delta_{d}sin(2\theta_{s})-i\Delta^{'}_{d}cos(2\theta_{s})}
\end{eqnarray*}

Further, for the partial derivatives of the scattering amplitudes appearing in Eq.~11, we have 
\begin{equation}
\partial_{V_{1}}S_{ee}=-\frac{\Delta^{'}_{d}cos(2\theta_{s})}{k(\Delta_{d}sin(2\theta_{s})-i\Delta^{'}_{d}cos(2\theta_{s}))}\mbox { and }
\partial_{V_{2}}S_{ee}=-\frac{\Delta^{'}_{d}cos(2\theta_{s})e^{-4ika}}{k(\Delta_{d}sin(2\theta_{s})+i\Delta^{'}_{d}cos(2\theta_{s})}
\end{equation}
and thus the pumped current into the normal metal lead for this order parameter becomes,
\begin{eqnarray}
I(NDd^{'})&=&\frac{-wq_{e}sin(\phi)V^{2}_{p}}{\pi k^{2}} \frac{\Delta^{'2}_{d}cos^{2}(2\theta_{s})}{\Delta^{2}_{d}sin^{2}(2\theta_{s})+\Delta^{'2}_{d}cos^{2}(2\theta_{s})} sin (4ka)
\end{eqnarray}

Furthermore, the ratio of the pumped current in presence of the High-T$_{c}$
superconductor with  pairing symmetry of the type $d_{x^{2}-y^{2}}+id_{xy}$ to that in a pure normal metal double barrier structure
(see Eq.~13) is

\begin{equation}
\frac{I(NDd^{'})}{I(N)}=4\frac{\Delta^{'2}_{d}cos^{2}(2\theta_{s})}{\Delta^{2}_{d}sin^{2}(2\theta_{s})+\Delta^{'2}_{d}cos^{2}(2\theta_{s})}
\end{equation}

 From the above expression it is evident that the maximum enhancement of the pumped current is $ 4$ times of that in a pure normal metal structure. Depending on the relative magnitudes of $\Delta^{'}_{d}$ and  $\Delta_{d}$ and ofcourse also depending on the injection angle the ratio  ${I(NDd^{'})}/{I(N)}$ can be as low as zero as in the pure $d_{x^{2}-y^{2}}$ case or as large as $4$ as in the pure $s-wave$ case.
 
 To conclude this section we have seen contrasting results in all the
 four cases, while as seen before for the $s-wave$ case there is four
 fold enhancement as compared to the normal metal case, in case of a
 $d_{x^{2}-y^{2}}-wave$ superconductor there is no pumped current at
 all, and  for the case of  $d_{x^{2}-y^{2}}+is -wave$ and  $d_{x^{2}-y^{2}}+id_{xy} -wave$ superconductor the enhancement depends on the relative magnitude of the components as well as the injection angle. To probe the dependence of the injection angle and relative magnitudes of the different components in the cases where we have considered mixed pairing symmetry, we in  figure 2 plot the the ratio of the pumped current in presence of the High-T$_c$ superconductor as function of the injection angle for different ratios of the relative magnitudes where mixed pairing symmetry is considered alongwith the pure $s-wave$ and $d_{x^{2}-y^{2}}-wave$ cases.

\begin{figure*}[h]
\protect\centerline{\epsfxsize=7.5in\epsfbox{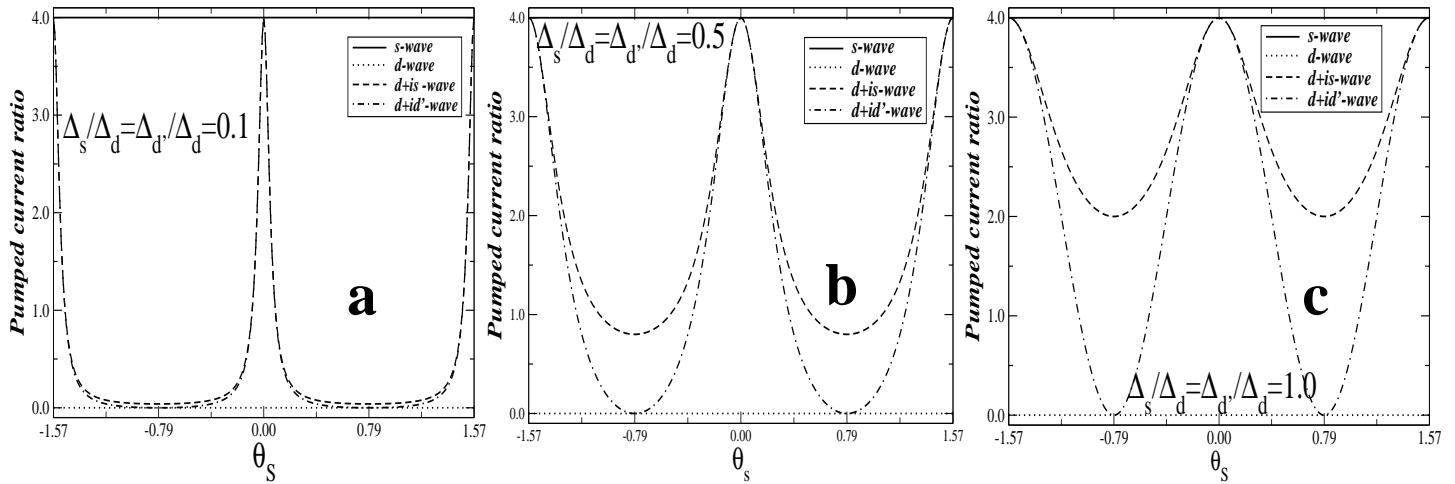}}
\caption{The ratio of the pumped current for double barrier quantum dot at resonance in junction with a High-$T_c$ superconductor to that in a pure normal metal double barrier structure. The panels:(a) The magnitude of the subdominant component in the mixed order parameter cases is $10\%$ of the dominant component, (b) The magnitude of the subdominant component in the mixed order parameter cases is $ half $ of the dominant component and (c)  The magnitude of the subdominant component in the mixed order parameter cases is $ equal $ to the dominant component.}
\end{figure*}

From the figure 2, it is quite evident that the $s-wave$ and $d_{x^{2}-y^{2}} -wave$ cases are completely independent of injection angle. Furthermore,
one can clearly see that whatever the strength of the subdominant component in the mixture i.e., $s$ or $d_{xy}$ for the injection angle $\theta_{s}=0,\pm \frac{\pi}{2}$, one has for the pumped current in $d_{x^{2}-y^{2}}+is -wave$ and  $d_{x^{2}-y^{2}}+id_{xy} -wave$ cases four times that in the pure normal metal structure. Also it is evident especially for the relative magnitudes of the sub-dominant component in the mixture being around half or more, that there is a marked difference between the $d_{x^{2}-y^{2}}+is -wave$ and  $d_{x^{2}-y^{2}}+id_{xy} -wave$ cases at injection angles $\theta_{s}=\pm \frac{\pi}{4}$, while for the $d_{x^{2}-y^{2}}+is -wave$ case the pumped current is almost same as that in a normal metal structure in figure 2(b), in figure 2(c) it is almost twice of that in a normal metal structure, but in both figures 2(b), and figure 2(c) the pumped current in the  $d_{x^{2}-y^{2}}+id_{xy} -wave$ case is zero at the same injection angle values. These differences can be easily exploited in distinguishing the different pairing symmetries considered here.

\section {Pumped heat and noise}
A time dependent scatterer always generates heat flows and can be
considered as a mesoscopic (phase coherent) heat source which can be
useful for studying various thermoelectric phenomena in mesoscopic
structures.  The adiabatic quantum pump thus not only generates an
electric current but also heat current which can be expressed as the
sum of noise power and the joule heat
dissipated\cite{buti_mosk_diss,wang_heat,wang_opt}. In this section we
look into the heat pumped and the noise generated for the various
order parameters of the High-T$_{c}$ superconductors considered above
to further unravel the differences among them.

The expressions for pumped heat and noise in the presence of a
superconducting ($s-wave$) lead have been earlier derived earlier in
Ref.[\onlinecite{wang_opt}]. Below we extend the description to
include the $d_{x^{2}-y^{2}}-wave$,  $d_{x^{2}-y^{2}}+is-wave$ and  $d_{x^{2}-y^{2}}+id_{xy} -wave$ superconductors.  The pumped current in Eq.~4, can be re-expressed as follows-

\begin{widetext}
\begin{equation}
I=\frac{wq}{2\pi}\int dE (-\partial_{E}f)\int_{0}^{\tau}dt \sum_{j=1,2} [\Im(S^{*}_{ee}\partial_{V_{j}}S_{ee}+S^{*}_{eh}\partial_{V_{j}}S_{eh})]\frac{d V_j}{d t}
\end{equation}
\end{widetext}
In the above equation, $\Im$ represents the imaginary part of the quantity in parenthesis. Furthermore, as in the adiabatic regime, $\partial_{t}
S_{\alpha\beta}=\sum_{i}[\partial_{V_i}S_{\alpha\beta}\partial_{t}X_{i}+...]$,
and from complex algebra
$\Im[S_{ee}^{*}\partial_{t}S_{ee}]=-i[S_{ee}^{*}\partial_{t}S_{ee}]$,
the pumped current becomes-

\begin{equation}
I=\frac{wq}{2\pi}\int dE \int_{0}^{\tau} dt [S^{}_{NS}\{f(E+i\frac{\partial_t}{2})-f(E)\}S^{\dagger}_{NS}]_{ee}
\end{equation}

with $S_{NS}$ being the $2X2$ matrix as defined in Eq.~1. In the
above equation the Fermi Dirac distribution is expanded to first order
in $\partial_t$ only and $[...]_{ee}$ represents the $ee^{th}$ element
of the quantity in brackets.

The heat current pumped is defined as the magnitude of the electric
current multiplied by energy measured from the Fermi level.
\begin{widetext}
\begin{equation}
H=\frac{1}{\pi\tau} \int_{0}^{\tau}\int dE (E-E_{F})[S_{NS}(E,t)\{f(E+i\frac{\partial_{t}}{2})-f(E)\}S^{\dagger}_{NS}(E,t)]_{ee}
\end{equation}
\end{widetext}

Expanding $f(E+i\frac{\partial_{t}}{2})$ up-to second order one gets a
non-vanishing contribution to the heat current in the zero temperature
limit as-

\begin{eqnarray}
H=\frac{1}{8\pi\tau}\int_{0}^{\tau}dt [\partial_{t}S_{NS}(E,t)\partial_{t}S^{\dagger}_{NS}(E,t)]_{ee}
\end{eqnarray}

and since two parameters are being varied, we have 

\begin{widetext}
\begin{equation}
H=\frac{1}{8\pi\tau}\int_{0}^{\tau}dt \sum_{i,j=1,2}[\partial_{V_i}S^{}_{ee}\partial_{V_j}S^{*}_{ee}+\partial_{V_i}S^{}_{eh}\partial_{V_j}S^{*}_{eh}]\frac{\partial V_{i}}{\partial t}\frac{\partial V_{j}}{\partial t}
\end{equation}

By integrating the above expression up-to $\tau=2\pi$ we get the pumped
current in the weak pumping regime as:

\begin{equation}
H=\frac{w^{2}V^{2}_{p}}{16\pi}[\sum_{\beta=e,h}|\partial_{V_1}S_{e \beta }|^{2}+\sum_{\beta=e,h}|\partial_{V_2}S_{e \beta }|^{2}+2cos(\phi)\sum_{\beta=e,h}\Re(\partial_{V_1}S_{e \beta}\partial_{V_2}S^{*}_{e \beta})]
\end{equation}

$\Re$ refers to the real part of the quantity in parenthesis.
Similar to the above one can derive expressions for the noise and
joule heat dissipated. The expression for the heat current can be
re-expressed as -

\begin{eqnarray}
H&=&\frac{1}{8\pi\tau}\int_{0}^{\tau}dt [\partial_{t} S_{NS}(E,t) S^{\dagger}_{NS}(E,t) S_{NS}(E,t) \partial_{t}  S^{\dagger}_{NS}(E,t)]_{ee}\nonumber\\
&=&\frac{1}{8\pi\tau}\int_{0}^{\tau}dt \sum_{\beta=e,h} [\partial_{t} S_{NS}(E,t) S^{\dagger}_{NS}(E,t)]_{e\beta}[ S_{NS}(E,t) \partial_{t}  S^{\dagger}_{NS}(E,t)]_{\beta e}
\end{eqnarray}

The diagonal term is identified as the joule heat while the
off-diagonal element is the noise power\cite{wang_opt}.

\begin{eqnarray}
H&=&J+N,\nonumber\\
&=& \frac{1}{8\pi\tau}\int_{0}^{\tau}dt [\partial_{t} S_{NS}(E,t) S^{\dagger}_{NS}(E,t)]_{ee}[ S_{NS}(E,t)  \partial_{t}  S^{\dagger}_{NS}(E,t)]_{ee} \nonumber\\
&+& \frac{1}{8\pi\tau}\int_{0}^{\tau}dt [\partial_{t} S_{NS}(E,t) S^{\dagger}_{NS}(E,t)]_{eh}[ S_{NS}(E,t)  \partial_{t}  S^{\dagger}_{NS}(E,t)]_{he}
\end{eqnarray}

Similar to the analysis for the pumped heat current, the joule heat
dissipated and the noise power can be expressed in the weak pumping
regime as

\begin{eqnarray}
J&=&\frac{V^{2}_{p}w^2}{16\pi}[\{\sum_{\beta=e,h}|S_{e\beta}\partial_{V_1}S_{e\beta}|^{2}+2\Re(S^{*}_{ee}S_{eh}\partial_{V_1}S_{ee}\partial_{V_1}S^{*}_{eh})\}
+\{\sum_{\beta=e,h}|S_{e\beta}\partial_{V_2}S_{e\beta}|^{2}+2\Re(S^{*}_{ee}S_{eh}\partial_{V_2}S_{ee}\partial_{V_2}S^{*}_{eh})\}\nonumber\\
&+ &2cos(\phi)\{\sum_{\beta=e,h}|S_{e \beta}|^{2} \Re (\partial_{V_1}S_{e \beta}\partial_{V_2}S^{*}_{e \beta})+\Re(S^{*}_{eh}S_{ee}\partial_{V_1}S^{}_{eh}\partial_{V_2}S^{*}_{ee})+\Re(S^{*}_{ee}S_{eh}\partial_{V_1}S^{}_{ee}\partial_{V_2}S^{*}_{eh})\}]
\end{eqnarray}

while the noise power is given as below:

\begin{eqnarray}
N&=&\frac{V^{2}_{p}w^2}{16\pi}[\{\sum_{\beta=e,h}|S_{h\beta}\partial_{V_1}S_{e\beta}|^{2}+2\Re(S^{*}_{he}S_{hh}\partial_{V_1}S_{ee}\partial_{V_1}S^{*}_{eh})\}
+\{\sum_{\beta=e,h}|S_{h\beta}\partial_{V_2}S_{e\beta}|^{2}+2\Re(S^{*}_{he}S_{hh}\partial_{V_2}S_{ee}\partial_{V_2}S^{*}_{eh})\}\nonumber\\
&+ &cos(\phi)\{\sum_{\beta=e,h}|S_{h \beta}|^{2} \Re (\partial_{V_1}S_{e \beta}\partial_{V_2}S^{*}_{e \beta})+\Re(S^{*}_{hh}S_{he}\partial_{V_1}S^{}_{eh}\partial_{V_2}S^{*}_{ee})+\Re(S^{*}_{he}S_{hh}\partial_{V_1}S^{}_{ee}\partial_{V_2}S^{*}_{eh})\}]
\end{eqnarray}

Now for our considered system, i.e., a double barrier quantum dot at
resonance, we have seen in the previous section that
$\partial_{V_j}S_{he}=\partial_{V_j}S_{eh}=0$ regardless of the order
parameter symmetry of the High-T$_c$ superconductor and hence the
expressions for the pumped heat, noise and joule heat dissipated
reduce to-

\begin{eqnarray}
H&=&\frac{V^{2}_{p}w^{2}}{16\pi}[|\partial_{V_1}S_{e e}|^{2}+|\partial_{V_2}S_{e e}|^{2}+2cos(\phi)\Re(\partial_{V_1}S_{e e}\partial_{V_2}S^{*}_{e e})]\\
J&=&\frac{V^{2}_{p}w^{2}}{16\pi}|S_{ee}|^{2}[|\partial_{V_1}S_{e e}|^{2}+|\partial_{V_2}S_{e e}|^{2}+2cos(\phi)\Re(\partial_{V_1}S_{e e}\partial_{V_2}S^{*}_{e e})]\\
N&=&\frac{V^{2}_{p}w^{2}}{16\pi}|S_{he}|^{2}[|\partial_{V_1}S_{e e}|^{2}+|\partial_{V_2}S_{e e}|^{2}+2cos(\phi)\Re(\partial_{V_1}S_{e e}\partial_{V_2}S^{*}_{e e})]
\end{eqnarray}
\end{widetext}
For our chosen system, i.e., the double barrier quantum dot at resonance in junction with the High-T$_c$ superconductor when $\epsilon=0$, we have, $|S_{ee}|^2=0$ and $|S_{he}|^{2}=1$, therefore, $J=0$ and $H=N$.
 
Now analyzing the above expressions for the different order
parameters, we have-

{\underline {$s-wave\mbox{ } superconductor$:}} In the $s-wave$ case
as we have already seen
$\partial_{V_1}S_{ee}=2\partial_{V_1}S_{11}=-i/k$ and
$\partial_{V_2}S_{ee}=2\partial_{V_2}S_{11}=-(i/k)(S_{12})^2$. With
this, the expression for the heat current pumped which is equal to the noise power  reduces to-

\begin{equation}
H=N=\frac{V^{2}_{p}w^{2}}{8\pi k^2}[1+cos(\phi)cos(4ka)].
\end{equation}

Thus as is evident from the expression for the pumped noise, the
quantum pump is non-optimal\cite{avron} (or, non-noiseless), only in
the special case when $4ka=(2n+1)\pi$ and $\phi=2n\pi$, with $ n=0,1,...$ 
is the optimality condition met. Of-course, $\phi=2n\pi$ implies that
in this case there is no charge current as well.

{\underline {$d_{x^{2}-y^{2}}-wave \mbox{ } superconductor$}:} In this
case as also seen earlier, we have $\partial_{V_1}S_{ee}=0$ and
$\partial_{V_2}S_{ee}=0$. Thus there is no heat pumped neither any
noise generated nor any joule heat dissipated. Thus the pump in
conjunct with a $d_{x^{2}-y^{2}}$-wave superconductor is cent-percent
optimal for any configuration of the parameters and under any
condition.

{\underline {$d_{x^{2}-y^{2}}+is-wave \mbox{ } superconductor$}:} From the previous section, we have  $\partial_{V_1}S_{ee}$ and $\partial_{V_2}S_{ee}$ for the order parameter in this case, with this the pumped
heat which is same as the noise generated in the pumping process reduces
to:

\begin{eqnarray*}
H&=&N=\frac{V^{2}_{p} w^{2}}{8\pi k^2}\frac{\Delta^{2}_{s}}{\Delta^{2}_{d}sin^{2}(2\theta_{s})+\Delta^{2}_{s}}[1+cos(\phi)cos(4ka)]
\end{eqnarray*}

Denoting the noise generated in the $s-wave$ case by $N_0$, we have the noise generated for this case becoming  just $N_{0}\Delta^{2}_{s}/(\Delta^{2}_{d}sin^{2}(2\theta_{s})+\Delta^{2}_{s})$.

{\underline {$d_{x^{2}-y^{2}}+id_{xy}-wave \mbox{ } superconductor$}:} From the previous section, we have  $\partial_{V_1}S_{ee}$ and $\partial_{V_2}S_{ee}$ for this case too, with this the pumped
heat which is same as the noise generated in the pumping process reduces
to:

\begin{eqnarray*}
H&=&N=\frac{V^{2}_{p} w^{2}}{8\pi k^2}\frac{\Delta^{'2}_{d}cos^{2}(2\theta_{s})}{\Delta^{2}_{d}sin^{2}(2\theta_{s})+\Delta^{'2}_{d}cos^{2}(2\theta_{s})}[1+cos(\phi)cos(4ka)]
\end{eqnarray*}

Denoting the noise generated in the $s-wave$ case by $N_0$, we have the noise generated for this case becoming  just $N_{0}\Delta^{'2}_{d}cos^{2}(2\theta_{s})/(\Delta^{2}_{d}sin^{2}(2\theta_{s})+\Delta^{'2}_{d}cos^{2}(2\theta_{s}))$.

To end this section we have seen that the pumped heat and noise
generated in the pumping process can also show marked differences for
the various order parameters considered. In the $s-wave$, the
$d_{x^{2}-y^{2}}+is -wave$ and the $d_{x^{2}-y^{2}}+id_{xy}-wave$  cases  the system is non-optimal while in
the $d_{x^{2}-y^{2}}- wave$ case it is cent percent optimal. Further
more in  the $d_{x^{2}-y^{2}}+id_{xy}-wave$ case the
pump may be turned optimal in some special situations as seen in Figure 3. These situations would help in differentiating between the order parameters for the mixed parameter cases. Especially for Figures 3(b) and 3(c) as it is quite clear that the pump is optimal (or,noiseless) in case of the  $d_{x^{2}-y^{2}}+id_{xy}-wave$ superconductor at injection angles $\theta_{s}=\pm \pi/4$.

\begin{figure*}[h]
  \protect\centerline{\epsfxsize=7.5in\epsfbox{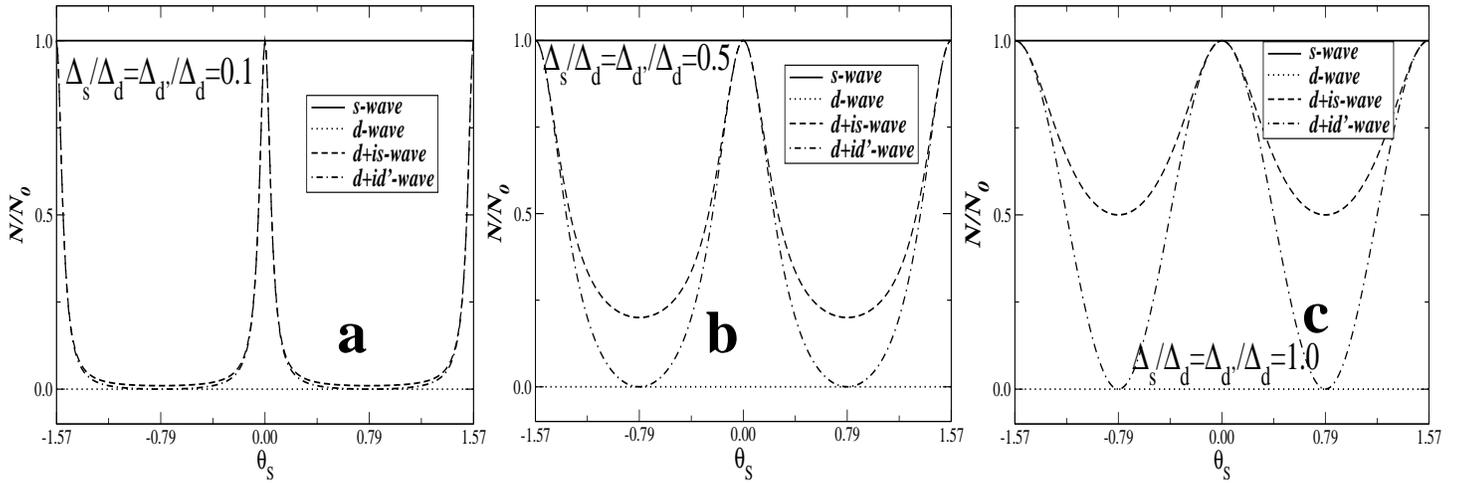}}
\caption{The noise generated for double barrier quantum dot at resonance in junction with a High T$_c$ superconductor. $N_{0}$ denotes the noise generated for the $s-wave$ case. The panels:(a) The magnitude of the subdominant component in the mixed order parameter cases is $10\%$ of the dominant component, (b) The magnitude of the subdominant component in the mixed order parameter cases is $ half $ of the dominant component and (c)  The magnitude of the subdominant component in the mixed order parameter cases is $ equal $ to the dominant component.}
\end{figure*}

\section {Experimental Realization}
 Although theoretical examples in quantum pumping phenomena are quite abundant, experiments in this field are very much lacking. Till date there have been notably four experiments in Refs.[\onlinecite{switkes,carlo,lorke}]and a quantum spin pump in Ref.[\onlinecite{spin_wu_wang}(b)]. The last reference deal with a quantum dot which with application of an inplane magnetic
 field  can pump a pure spin current. One can suitably modify these experiments and place the quantum dot in junction with a High-T$_c$ superconductor. The resonant condition of the quantum dot can be easily established by applying a suitable gate voltage which will enable resonant transport through the quantum dot. After this the two delta barrier's can be two gates which control the charge on the dot,  modulating these two gates in time will enable a pumped charge (also heat and noise) current to flow. This set-up can easily establish the results arrived at in this work and hopefully give more clues in building a correct theory for High T$_c$ superconductors.

\begin{table*}[!]
\begin{center}
\begin{tabular}{|c|c|c|c|c|}
\hline
   Order Parameter$\rightarrow$& $s-wave$&  $d_{x^{2}-y^{2}}-wave$  & $d_{x^{2}-y^{2}}+is -wave$  &  $d_{x^{2}-y^{2}}+id_{xy} -wave$ \\ \hline
  Pumped$\downarrow $     &            &           &       &      \\ \hline
Charge&$\frac{I(NS)}{I(N)}=4$&0&$\frac{I(NDs)}{I(N)}=\frac{\Delta^{2}_{s}}{\Delta^{2}_{d}sin^{2}(2\theta_{s})+\Delta^{2}_{s}}$& $\frac{I(NDd')}{I(N)}= \frac{\Delta^{'2}_{d}cos^{2}(2\theta_{s})}{\Delta^{2}_{d}sin^{2}(2\theta_{s})+\Delta^{'2}_{d}cos^{2}(2\theta_{s})}$\\ \hline
Heat&$H_{0}$&0&$H_{0}\frac{\Delta^{2}_{s}}{\Delta^{2}_{d}sin^{2}(2\theta_{s})+\Delta^{2}_{s}}$  &$H_{0} \frac{\Delta^{'2}_{d}cos^{2}(2\theta_{s})}{\Delta^{2}_{d}sin^{2}(2\theta_{s})+\Delta^{'2}_{d}cos^{2}(2\theta_{s})}$ \\ \hline
 Noise &\em Non-Optimal$^{}$ & \em cent percent Optimal & \em Non-Optimal$^{}$ &\em Non-Optimal$^{*}$ \\ \hline
\end{tabular}
\vskip 0.25in
\caption{\small{A comparative analysis of pumped charge, heat and noise in cases of $s-wave$, $d_{x^{2}-y^{2}}-wave$, $d_{x^{2}-y^{2}}+is -wave$ and  $d_{x^{2}-y^{2}}+id_{xy} -wave$ superconductors in conjunct with a normal metal double barrier structure. [$H_{0}=\frac{V^{2}_{p}w^{2}}{8\pi k^2}[1+cos(\phi)cos(4ka)]$, $*$ Optimal for injection angles $\theta_{s}=\pm \pi/4 $ (see section Pumped Heat and Noise)].}}\vspace{-9mm}
\end{center}
\end{table*}
\section {Conclusions}

To conclude we have given a simple procedure to distinguish various
order parameters proposed in the context of High-T$_{c}$
superconductivity. In Table 1 above we juxtapose the results
obtained in this work. The pumped charge current, heat pumped and
noise generated for the four cases considered, that of the $s-wave$,
$d_{x^{2}-y^{2}}- wave$,  $d_{x^{2}-y^{2}}+is -wave$ and  $d_{x^{2}-y^{2}}+id_{xy} -wave$ vary markedly which easily reveals the differences among the three.


\begin{thebibliography}{99}

\bibitem{bruder} C. Bruder, \prb {\bf 41}, 4017 (1990); C. Bruder,
  G. Blatter, and T. M. Rice, Phys. Rev. B {\bf 42}, R4812 (1990).

\bibitem{rice} M. Sigrist and T. M. Rice, \rmp {\bf 63}, 239 (1991).

\bibitem{vanhar} D. J. van Haarlingen, \rmp {\bf 67}, 515 (1995).

\bibitem{tseui} C. C. Tsuei and J. R. Kirtley, \rmp {\bf 72}, 969 (2000).

\bibitem{stefankis} N. Stefanakis and N. Flytzanis, \prb {\bf 64},
  024527 (2001); N. Stefanakis, \prb {64}, 224502 (2001).

\bibitem{asano1} Y. Asano, Y. Tanaka and S. Kashiwaya, \prb {\bf 69},
  134501 (2004).

\bibitem{asano2} Y. Asano, Y. Tanaka and S. Kashiwaya,
  cond-mat/0302287.

\bibitem{shiba} M. Matsumoto and H. Shiba, J. Phys. Soc. Japan {\bf
    64}, 3384 (1995).

\bibitem{kirtley} J. R. Kirtley, et. al., \prb {\bf 51}, R12057 (1995).

\bibitem{hu} C-R. Hu, \prl {\bf 72}, 1526 (1994); J. Yang and C-R. Hu, 
  \prb {\bf 50}, R16766 (1994).

\bibitem{tanaka} Y. Tanaka and S. Kashiwaya, \prl {\bf 74}, 3451
  (1995).

\bibitem{becherer} Th. Becherer, C. Stolzel, G. Adrian and H. Adrian, \prb {\bf 47}, R14650 (1993);
  T. Walsh, \ijmp {\bf 6}, 125 (1992) and references therein.

\bibitem{zhu} Jian-Xin Zhu and C. S. Ting, \prb {\bf 59}, R14165
  (1999).


\bibitem{belogovskii} M. Belogolovskii, \prb {\bf 67}, 100503(R)
  (2003); M. Belogolovskii, et. al., \prb {\bf 59}, 9617 (1999).

\bibitem{zutic} I. Zutic and O. T. Valls, \prb {\bf 61}, 1555 (2000);
  I. Zutic, J. Fabian and S. Das Sarma, \rmp {\bf 76}, 323 (2004).

\bibitem{hirai} T. Hirai, et. al., cond-mat/0210693.

\bibitem{kashiwaya}  S. Kashiwaya, Y. Tanaka, N. Yoshida and
  M. R. Beasley, \prb {\bf 60}, 3572 (1999).

\bibitem{deutscher} G. Deutscher, cond-mat/0409225.

\bibitem{lofwander} T. L\"{o}fwander, V. S. Shumeiko and G. Wendin,
  cond-mat/0102276. 
  
\bibitem{varma} Tai-Kai Ng and C. M. Varma, Phys. Rev. B {\bf 70},
  054514 (2004).

\bibitem{brouwer} P. W. Brouwer, \prb {\bf 58}, R10135 (1998).

\bibitem{zhou} F. Zhou, B. Spivak and B. Altshuler, \prl {\bf 82},
  608 (1999).

\bibitem{bpt} M. B\"{u}ttiker, H. Thomas and A. Pretre, Z. Phys. B:
  Condensed Matter {\bf 94}, 133 (1994).

\bibitem{switkes} M. Switkes, C. M. Marcus, K. Campman and A.
  C. Gossard, Science {\bf283}, 1905 (1999); M. Switkes, Ph. D thesis,
  Stanford University (1999).
  
\bibitem{spin_wu_wang} E. R. Mucciolo, C. Chamon and C.  M.  Marcus,
  \prl {\bf 89}, 146802, (2002); S.  K. Watson, R. M. Potok, C. M.
  Marcus and V.  Umansky, Phys. Rev. Lett {\bf 91}, 258301 (2003); J.
  Wu, B. Wang, and J. Wang, \prb {\bf 66}, 205327 (2002); W. Zheng,
  et.al, Phys. Rev. B {\bf 68}, 113306 (2003); P. Sharma and C.
  Chamon, \prl {\bf 87}, 096401 (2001); T.  Aono, \prb {\bf 67},
  155303 (2003); M.  Governale, F.  Taddei and R. Fazio, \prb {\bf
    68}, 155324 (2003); P.  Sharma and P. W. Brouwer, Phys. Rev.
  Lett. {\bf 91}, 166801 (2003); R. Benjamin and C. Benjamin, \prb
  {\bf 69}, 085318 (2004); Yadong Wei, L. Wan, B. Wang and J. Wang,
  \prb {\bf 70}, 045418 (2004); K. Yu Bliokh and Yu P Bliokh,
  quant-ph/0404144; F. Zhou, \prb {\bf 70}, 125321 (2004); A. Brataas
  and Y. Tserkovnyak Phys.  Rev. Lett. {\bf 93}, 087201 (2004); M.
  Yang and S-S. Lee, \prb (to be published); E.  Sela and Y. Oreg,
  cond-mat/0407089; C. Bena and L. Balents, cond-mat/0406437; Y.
  Tserkovnyak, A.  Brataas, G.  E.  W. Bauer, and B. I. Halperin,
  cond-mat/0409242.
  
\bibitem{blau_1} M. Blaauboer, \prb {\bf 68}, 205316 (2003);
  cond-mat/0307166.

\bibitem{citro} R. Citro, N. Andrei and Q. Niu, \prb {\bf 68}, 165312
  (2003); S. Das and S. Rao, cond-mat/0410025.

\bibitem{aharony} Amnon Aharony and O. Entin-Wohlman, Phys. Rev. B
  {\bf 65}, 241401(R) (2002);  V. Kashcheyevs, M. Sc. thesis (2002).

\bibitem{wang_kondo} B. Wang and J. Wang, \prb {\bf 65}, 233315 (2002).

\bibitem{wang_apl} Jian Wang, Yadong Wei, Baigeng Wang and Hong Guo,
  \apl {\bf 79}, 3977 (2001).

\bibitem{blaau} M. Blaauboer, \prb {\bf 65}, 235318 (2002).
  
\bibitem{taddei} F. Taddei, M. Governale and R. Fazio, \prb {\bf 70}
  052510 (2004).

  
\bibitem{andreev} A. F. Andreev, Sov. Phys. JETP {\bf 19}, 1228
  (1964); P. De-Gennes and D. S. James, Phys. Letters {\bf 4}, 151
  (1963).

\bibitem{beenakker_rmp} C. W. J. Beenakker, \rmp {\bf 69}, 731 (1997).

\bibitem{beenakker} C. W. J. Beenakker, \prb {\bf 46}, R12841 (1992).

\bibitem{wang_comm} see citation 27 in B. Wang and J. Wang, \prb {\bf
    66} 201305(R) (2002); Jian Wang, Private communication (2004).

\bibitem{gasparian} V. Gasparian, T. Christen and M. B\"{u}ttiker,
  \pra {\bf 54}, 4022 (1996).

\bibitem{fisher-lee} D. S. Fisher and P. A. Lee, \prb {\bf 23}, R6851
  (1981).

\bibitem{wang_heat}  B. Wang and J. Wang, \prb {\bf 66}, 125310
  (2002).

\bibitem{wang_opt} B. Wang and J. Wang, \prb {\bf 66}, 201305(R)
  (2002).
  
\bibitem{buti_mosk_diss} M. Moskalets and M. B\"{u}ttiker, Phys. Rev.
  B {\bf66}, 035306 (2002); M. Moskalets and M. B\"{u}ttiker,
  cond-mat/0407292.
  
\bibitem{avron} J.E. Avron, A. Elgart, G.M. Graf, and L. Sadun, \prl
  {\bf 87}, 236601 (2000); J.E. Avron, A. Elgart, G.M. Graf, and L.
  Sadun, math-ph/0305049; Kai Schnee, Dissertation (2002).

\bibitem{carlo} L. DiCarlo, C. M. Marcus and J. S. Harris, Jr., \prl {\bf 91}, 246804 (2003).

\bibitem{lorke} E. M. H\"{o}hberger, A. Lorke, W. Wegscheider and M. Bichler,
 \apl {\bf 78}, 2905 (2001).
\end{thebibliography}
\end{document}